\documentclass[10pt,twocolumn,twoside]{IEEEtran}

\usepackage{graphicx}
\usepackage{multirow}
\usepackage{booktabs}
\usepackage{mathtools}
\usepackage{tabu}
\usepackage{bm}
\usepackage[numbers,sort&compress]{natbib}
\usepackage{epstopdf}

\begin{document}

\title{Efficient Multiple Line-Based Intra  Prediction\\ for HEVC }

\author{Jiahao Li,
	Bin Li, \IEEEmembership{Member,~IEEE,}
	Jizheng Xu,\IEEEmembership{ Senior Member,~IEEE,}
	and  Ruiqin Xiong, \IEEEmembership{Member,~IEEE}
	\thanks{
		Manuscript received Aug. 19, 2016; revised Oct. 31, 2016; accepted Nov.
		22, 2016.
		This work was supported in part by the National Basic Research Program of China under Grant 2015CB351800, the National Natural Science Foundation of China under Grant 61370114, 61425026 and 61421062, and also by the Cooperative Medianet Innovation Center.
	}
    \thanks{This work was done when J. Li was with Microsoft Research Asia.}
	\thanks{J. Li and R. Xiong are with the Institute of Digital Media, School of Electronic Engineering and Computer Science, and the National Engineering Laboratory for Video Technology (NELVT), Peking University, Beijing 100871, China (e-mail: jhli.cn@pku.edu.cn, rqxiong@pku.edu.cn).}
	\thanks{B. Li and J. Xu are with Microsoft Research Asia, Beijing 100080, China (e-mail: libin@microsoft.com, jzxu@microsoft.com).}
	\thanks{Copyright © 2016 IEEE. Personal use of this material is permitted. However, permission to use this material for any other purposes must be obtained from the IEEE by sending an email to pubs-permissions@ieee.org.}

}

%\markboth{Submitted to IEEE Transactions on Circuits and Systems	for Video Technology}
%{Li \MakeLowercase{\textit{et al.}}: Efficient Multiple Line-Based Intra Prediction for HEVC }

\maketitle

\begin{abstract}
Traditional intra prediction usually utilizes  the nearest reference line to generate the predicted block when  considering  strong spatial correlation. However, this kind of single  line-based method does not always work well due to at least two issues.
	One is the incoherence  caused by the signal noise or the texture of other object, where this texture deviates from the inherent texture of the current block.
	The other reason is that the nearest reference line  usually has worse reconstruction quality in  block-based video coding.
	Due to these two issues, 	this paper proposes an efficient multiple line-based intra prediction scheme to improve  coding efficiency.  Besides the nearest reference line,  further reference lines are also utilized. The  further reference lines with relatively higher quality can  provide potential  better prediction.
	 At the same time, the residue compensation is  introduced to calibrate the prediction of  boundary regions in a block when we utilize  further reference lines. 
	  To speed up the encoding process, this paper  designs several fast algorithms.
	   Experimental results show that, compared with HM-16.9, the proposed fast search method achieves $2.0\%$ bit saving on average and up to $3.7\%$, with increasing the encoding time by $112\%$.
\end{abstract}


\IEEEpeerreviewmaketitle

\section{Introduction}
\IEEEPARstart{I}{ntra} coding is important for  video coding to explore the spatial correlation. In the newly published High Efficiency Video Coding  \cite{sullivan2012overview} standard, many new technologies      have been introduced for  intra coding. These improvements have helped HEVC become a state-of-the-art video compression scheme, which can provide a similar perceptual quality with  about $50\%$ bitrate saving  compared with its predecessor H.264/AVC \cite{ohm2012comparison}.

In HEVC, the number of intra  modes has been extended to $35$ which includes planar mode, DC mode, and $33$ angular modes \cite{lainema2012intra}. This kind of fine-grained modes can provide more accurate prediction by leading to lower residual power when compared with the intra prediction in H.264/AVC, in which there are only  $9$  modes \cite{chen2015new}. 
In H.264/AVC, the macro block size is fixed on $16\times16$. Only $4\times4$, $8\times8$, and $16\times16$  intra predictions  are allowed. By contrast, HEVC uses a more flexible quadtree structure to choose the best block size adaptively, with the newly introduced     concepts of coding unit (CU), prediction unit (PU), and transform unit (TU) \cite{HEVCdraft}. PU carries  prediction information such as intra direction, and varies from $4\times4$ to $64\times64$. TU is the basic unit for intra prediction, and varies from $4\times4$ to $32\times32$. All TUs in a PU will share the same prediction direction.  As the partitions of PU and TU provide the flexibility to adapt the content,  it improves  intra coding efficiency. In addition, HEVC introduces the adaptive scanning order and discrete sine transform (DST) to further improve  intra coding efficiency.

    \begin{figure}
    	\centering
    	\includegraphics[width=260pt, height=140pt]{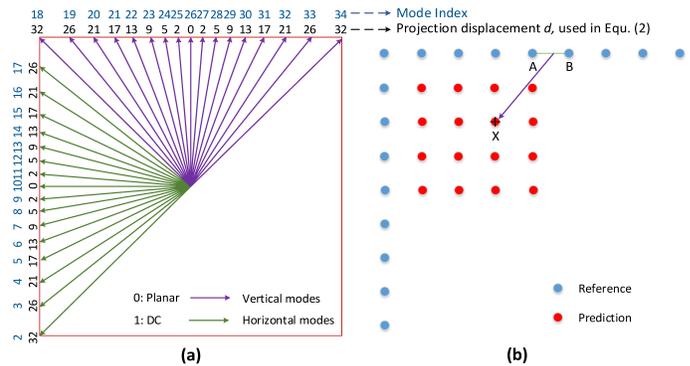}\\
    	\caption{ HEVC intra prediction. (a) 35 modes.  (b) Angular prediction illustration. }\label{HEVCIntra}
    \end{figure}
Fig. \ref{HEVCIntra} shows the intra prediction in  HEVC. For angular prediction, each pixel is projected to the nearest reference line along the angular direction. A two-tap linear interpolation filter with   1/32 pixel accuracy   is used to generate  the prediction,  where the filter coefficient is the inverse proportion of the two distances between the projected fraction position and its two adjacent integer positions. 
For  DC mode, the average of the pixels in the nearest reference line is as the predictor. Bi-linear interpolation is used in planar mode.

There are many  works that further improve  intra prediction. To better explore  spatial correlation, many works \cite{kamisli2012intra,kamisli2013intra,kamisli2015block,chen2013recursive,li2014rate}  model  the correlation between adjacent pixels as a first order 2-D markov process,  where each pixel is predicted by linearly weighing several adjacent pixels. The weightings in the Markov model need to be trained off-line. 
Another typical technology used for improving intra prediction is  image inpainting, and it can be subdivided into edge-based \cite{liu2007image,liu2008edge}, partial differential equations-based \cite{doshkov2010towards,zhang2014improving}, and total-variation based \cite{qi2012intra} methods. In addition,  \textit{Lai et al.} propose an error diffused intra prediction algorithm for HEVC \cite{lai2015error}.  \textit{Zhang et al.} propose a position-dependent filtering method \cite{zhang2011novel}. In \cite{chen2016improving}, an iteratively filtering-based scheme is proposed. 
The weighed prediction with two  directions has also been investigated in \cite{BiIntra,ye2008improved,yeh2015predictive}.
 Recently, \textit{Chen et al.} have proposed  to encode one half of the pixels in a block first, and the remains are reconstructed   by interpolations \cite{chen2015new}. The coding gain mainly comes from  the prediction distance being shortened.

However, these aforementioned works  and the intra prediction in the HEVC standard 　only utilize the nearest reference line to generate the prediction. Actually some further regions   can  also be utilized as there may exist similar content which can be used as the predictor. One famous technique is the template matching \cite{tan2006intra, tan2007intra,guo2008priority,zhang2015hybrid} which searches similar non-local  content by using the neighbor reference pixels as an indicator. It performs well on the situation that there are many repeated patterns. As there is a searching procedure for both  encoder and decoder, it has a high level of complexity, especially for the decoder. Another similar technique is the intra block copy \cite{ibc}, which is commonly used in screen content coding. When compared with template matching, the major difference is that the predictor is directly indicated by displacement vector. This avoids the searching procedure at the decoder side.

In this paper, we  propose  taking advantage of the local further regions, namely the further reference lines. The utilization of further reference lines has been investigated in \cite{matsuo2009intra}.  However, our method is different from \cite{matsuo2009intra}, and the detailed comparison will be described in Section III. The work \cite{matsuo2009intra} was implemented on H.264/AVC reference software, and its method without gradient achieves $1.2\%$ gain on average, with increasing the encoding time by $1228\%$. The method with gradient in \cite{matsuo2009intra} achieves $2.0\%$ gain on average, with increasing the encoding time by $3836\%$. In addition, multiple line-based intra prediction was proposed for JEM (Joint Exploration Model)  \cite{MultiLine, MultiLineITRI}.

 Two main reasons motivate us to develop the multiple line-based intra prediction method. 
One is that the nearest reference line may have signal noise or the texture of other object, where  this texture is inconsistent with the inherent texture of the current block.
This will cause the incoherence between the nearest reference line and current block, and then decrease the prediction accuracy.
In such cases, using the further reference lines is helpful because more predictions are provided for choosing.
 The other reason is that  the pixels in different positions of a block have  different reconstruction quality. In most block-based video coding frameworks, the residues are quantized in the transform domain.  The unequal frequency quantization error in the  transform domain  will lead to  different spatial quantization errors in  the pixel domain \cite{robertson2005dct}.  In general, the boundary of a block has a larger quantization error (i.e. worse quality), especially at the corners. So the nearest reference line utilized by traditional intra prediction solutions is often with  worse quality. 

Because of these two reasons, this paper designs a multiple line-based intra prediction scheme to improve   coding efficiency. The main contributions of this paper are as follows:1)  A residue compensation procedure is introduced to  calibrate the prediction when further reference lines are used. 2) This paper  designs several fast encoding algorithms to control encoder complexity. 
3) Based on residue compensation and fast algorithms, we propose an efficient multiple line-based intra prediction scheme. 
 The predictions generated from different reference lines, including the traditional prediction generated from the nearest reference line, are selected by the rate distortion optimization to choose the best prediction for each  block. Experimental results verify  that the proposed algorithm is effective. For all intra coding, the bit saving of the proposed fast search scheme is  $2.0\%$  on average, and up to $3.7\%$, with  increasing the encoding time by $112\%$. The bit saving of full search scheme is  $2.3\%$  on average, and up to $4.3\%$. 

The rest of this paper is organized as follows. Section II reveals the motivation of the proposed method with detailed analyses.
Section III introduces the proposed method including the basic framework,  residue compensation, and corresponding fast algorithms.  Experimental results are presented in Section IV.  Section V concludes this paper.

\section{ Analysis of Intra Prediction } 
As introduced in the previous section, we are motivated by two reasons to develop the  multiple line-based intra prediction scheme. In this section, we will analyze them in detail.
First, we  analyze the incoherence caused by signal noise or the texture of other object.
Then we concretely reveal the unequal reconstruction quality  of different regions in a block, and its influence on intra prediction.

 \subsection{ Incoherence Between The Nearest Reference Line and The Current Block }
 \begin{figure}
 	\centering
 	\includegraphics[width=255pt, height=233pt]{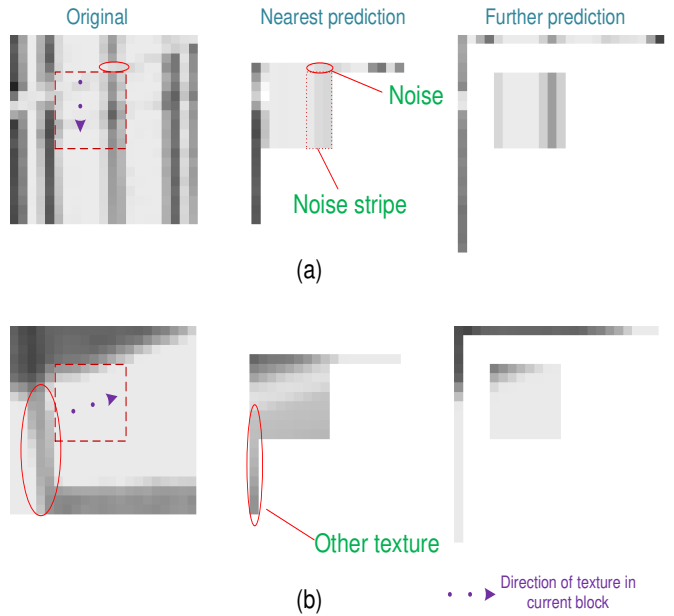}\\
 	
 	\caption{ The examples of possible cases. $8\times 8$ block. The left block is the original block. The middle block is the predicted block generated by the nearest reference line. The right block is the predicted block generated by the further reference line. (a) \textit{BQTerrace}, best direction of nearest  prediction: 26; best direction of further prediction: 26. (b) \textit{PeopleOnStreet}, best direction of nearest  line: 7; best direction of further line: 6. }\label{NoiseExample}
 \end{figure}
 In essence, the angular prediction in HEVC is a copying-based process with the assumption that  visual content follows a pure direction of propagation. Since the pixels with  shorter distance generally have  stronger  correlation in a picture,  HEVC only utilizes the nearest reference line to predict the current block. 
 However, the nearest reference line can not always predict well due to the incoherence, which is caused by signal noise or the texture of other object.
 During video acquisition, the noise is introduced.
If the nearest reference line is noised and used for prediction, the noise will also  be propagated into the whole predicted block, forming the  noise stripe. The noise stripe will make the prediction deviate from the original.
In addition, the  texture of other object may emerge on the nearest reference line,  where the  texture is inconsistent with the  inherent texture of the current block. This kind of other texture will break the prediction of inherent texture.
These two issues will cause the incoherence between the nearest reference line and current block.
 We  illustrate  two examples corresponding to the two  issues, as shown in Fig \ref{NoiseExample}. 
 Fig. \ref{NoiseExample} (a) shows an example of signal noise, and Fig. \ref{NoiseExample} (b) is an example with other texture.
 From the middle block, we can see the prediction generated by the nearest reference line has an obvious deviation from the left original block, which is not what we expect.

 For this reason, we expect to utilize the further reference lines to seek potential better prediction when there exists incoherence. This is because that  further reference lines may have less noise or may keep the   inherent texture of the current block.
 In the two aforementioned examples, the   further reference line can  provide  more precise prediction, as shown in the right block in Fig \ref{NoiseExample}.

\begin{figure}
 	\centering
 	\includegraphics[width=250pt, height=120pt]{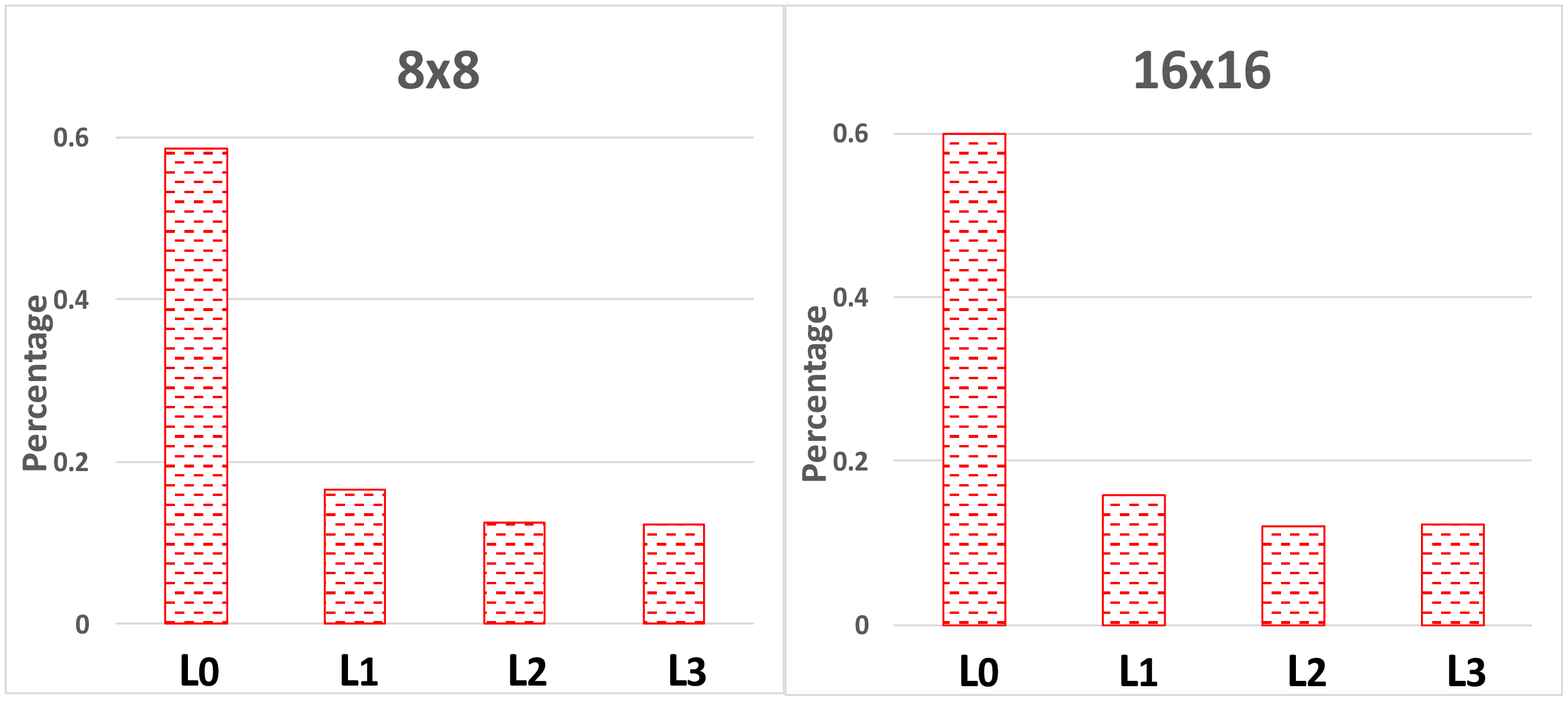}\\
 	\caption{ The percentage of each reference line.
 		The nearest reference line: $L_{0}$. The further  reference lines: $L_{1}$, $L_{2}$, and $L_{3}$.  }\label{AnaBetterRatio}
 \end{figure}

 This kind of incoherence is not uncommon in intra prediction, and there exist many cases that  further reference line can  provide better prediction.
 To verify our assumption, we collect the percentage number of each reference line  if using the multiple reference lines, as shown in Fig. \ref{AnaBetterRatio}. 
 The best reference line of each block is chosen from the nearest 4 reference lines according to the sum of absolute transformed differences (SATD) between the predicted block and the original block, where each reference line will check 35 directions just as HEVC intra prediction.
 The statistic is from the first picture of five sequences  (\textit{BasketballDrill, BQTerrace, Cactus, Traffic,}  and \textit{PeopleOnStreet}).
  Each picture will be partitioned into non-overlapped blocks with a fixed size for prediction,  where we investigate the block sizes $8\times8$ and $16\times16$ separately.  To exclude the influence of the quantization error in the reference line, the prediction is generated by the original pixels of the reference line rather than compressed pixels. 
 In this figure, although the nearest reference line ($L_{0}$) takes the largest percentage, there are still up to $41\%$ blocks  choosing further reference lines ($L_{1}$, $L_{2}$, and $L_{3}$) for $8\times 8$ blocks.
 This significant  percentage shows the potential benefits if utilizing further reference lines. Similarly, the percentage of choosing further reference lines in  $16\times 16$ blocks is $40\%$.

\subsection{Reconstruction Quality of References}
 \begin{figure}
	\centering
	\includegraphics[width=250pt, height=230pt]{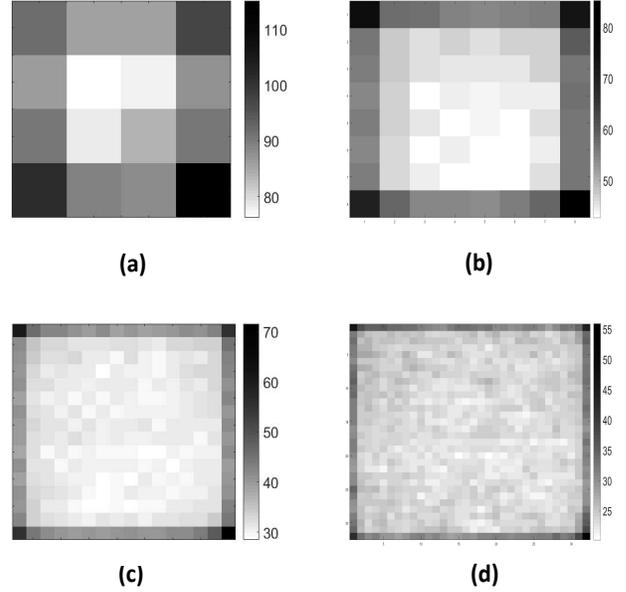}\\
	\caption{ The variance of spatial quantization error in HEVC intra coding, Luma component, quantization parameter: 37. The variance is calculated before the in-loop filters such as deblocking and sample adaptive offset are applied because the intra coding uses the  reconstruction without in-loop filters as the predictor. (a)$ \sim$(d):    $4\times4 \sim 32\times32$. }\label{SpatialError}
\end{figure}
In most video/image  coding frameworks, the  residue is organized as  block and quantized in the transform domain. The quantization  introduces the frequency quantization error. As the transform and inverse transform are  linear operations, the spatial quantization error in the pixel domain is the inverse transform of the frequency quantization error in the transform domain. Furthermore, from a statistical perspective, the relation between the variance of  spatial quantization error and that of frequency quantization error also can be derived theoretically.  \textit{Robertson et al.} \cite{robertson2005dct}  present the relation for the discreet cosine transform (DCT). They conclude that the locations near the block boundary have  relatively higher  spatial quantization error variance for smooth signals whose  signal energy is mainly contained in the low frequency coefficients. On the contrary, for signals that contain significant high-frequency content, such as textured regions, the inner pixels of a block  have  higher spatial quantization error variance.

In most cases, the  smooth blocks take a larger proportion in a sequence when compared with the complex texture blocks,  leading to  the error variance of the boundary region being larger than that of the inner region, according to the conclusion of \cite{robertson2005dct}. To verify the inference, we  calculate  the variance of the spatial quantization error using statistics.
Fig. \ref{SpatialError} shows the variance of spatial quantization error for different block sizes of five sequences (\textit{BasketballDrill, BQTerrace, Cactus, Traffic,}  and \textit{PeopleOnStreet}). These sequences are compressed by the HEVC reference software HM-16.9 with  all intra configuration, and the quantization parameter (QP) is set to 37.
 From the figure, we can see that the border  has the worst quality in a block, especially at the corners. However, the  right most column and the  bottom most row with the worst quality are exactly  the reference pixels used for traditional intra prediction. This deviates from our expectation  that the reference pixels used for prediction should be  high quality.    It is noted that the 4x4 luma block takes DST in HEVC intra coding, and blocks with other sizes use DCT. Nevertheless, these blocks follow  similar characteristics from the quality distributions shown in Fig. \ref{SpatialError}.

 Further reference lines with relatively better quality will bring more benefits for intra prediction. To verify this, we  collect the percentage numbers just like the analysis in  Fig. \ref{AnaBetterRatio}.  We merely generate the prediction with the compressed pixels of reference lines rather than original pixels,  where the QP is 37  for compression.
 We find that the percentage of prediction from further reference lines has increased from  $41\%$ to $60\%$ for the $8\times8$ block. The  increase of $19\%$ verifies that further reference lines with better quality have higher possibility of being  chosen.
Similarly, the percentage of the $16\times16$ block  increases  from  $40\%$ to $61\%$. In this subsection, we conclude that the further reference lines with relatively better quality have a stronger reason to be utilized in video coding.

\section{ Multiple Line-Based Intra Prediction}

Based on the analyses from the previous section, this paper proposes an efficient multiple  line-based scheme for HEVC. 
In this section, we  first introduce how to generate the predicted block with a further reference line, since it is the basis in the proposed method. In the procedure of predicting with further reference lines, some affiliated information  can also be obtained. To take the full advantage of this information, we propose a residue compensation as a post processing to further refine the predicted block.
 Besides  coding efficiency, we also care about the encoding complexity.
  For this reason, a fast solution is designed, into which several acceleration algorithms are incorporated.

\subsection{Prediction Generation}
\begin{figure}
	\centering
	\includegraphics[width=265pt, height=205pt]{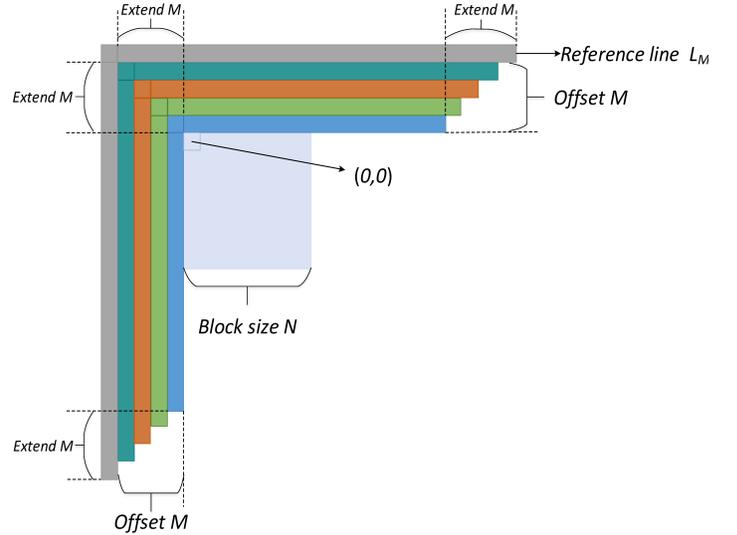}
	
	\caption{ The structure of the multiple reference lines.}\label{reference_line_structure}
\end{figure}
In the proposed scheme,  the structure of the multiple reference lines is organized  as shown in Fig.  \ref{reference_line_structure}. Each reference line is indicated by an index $L_{M}$ (the nearest reference line  corresponds to index $L_{0}$). From Fig. \ref{reference_line_structure}, we can see that there exists an interval between the reference line $L_{M}$ and the predicted block, whose offset is $M$.

In the generation of a predicted block, the further reference line takes the same rule as the nearest reference line,   which is specified in the HEVC standard \cite{lainema2012intra}.
 For the 33 angular directions, each pixel of predicted block is projected to  reference line $L_{M}$ along the direction,  then  the   interpolated  value (at 1/32 pixel accuracy) is used as the predictor:
 \begin{equation}\label{IntraGenerate1}
 	p_{x,y}=((32-z_{y})\cdot r_{i,-M-1}+z_{y}\cdot r_{i+1,-M-1} +16)>>5.
 \end{equation}
 $p_{x,y}$ is the predicted value of each pixel, where $x$ and $y$ are the column and row  index, respectively. $(0,0)$ represents the most top-left pixel in the current predicted block. $r_{x,y}$ is the  reconstructed value of pixel at $(x,y)$ position in other neighboring blocks which have already been coded.
 In particular,  $r_{i,-M-1}$ and $r_{i+1,-M-1}$ are the two adjacent  pixels corresponding to the projected subpixel in reference line $L_{M}$.  $>>$ denotes a bit shift operation to the right. Reference pixel index $i$ and interpolation parameter   $z_{y}$ are calculated according to position of current pixel and the projection displacement $d$ (a value from $-32$ to $+32$ as shown in Fig. \ref{HEVCIntra} (a)) which  is associated with current prediction direction.
  \begin{equation}\label{IntraGenerate2}
  \begin{split}
  &c_{y}=((y+M+1)\cdot d)>>5\\
  &z_{y}=((y+M+1)\cdot d)\&31\\
  &i=x+  c_{y},
  \end{split}
  \end{equation}
  where \& is bitwise AND operation.
  Equations (\ref{IntraGenerate1}) and (\ref{IntraGenerate2}) define the generation of prediction for vertical modes (i.e.  mode $18  \sim 34$). Meanwhile, the reference line $L_{M}$ is reorganized as a unified reference row (i.e. projecting the left reference column to extend the top reference row toward left, and more details can be found in \cite{lainema2012intra}). For horizontal modes (i.e.  mode $2  \sim 17$), the  prediction is derived identically by swapping the $x$ and $y$ coordinates in Equations (\ref{IntraGenerate1}) and (\ref{IntraGenerate2}), and the $L_{M}$ is reorganized as a unified reference column (i.e. projecting the top reference
  row to extend the left reference column upward).
  
For DC mode, the average of the  reference line $L_{M}$ is used as the predictor.
In planar mode, each pixel is a bi-linear interpolation:
 \begin{equation}\label{IntraGenerate3}
 \begin{split}
 &p^{V}_{x,y}=((N-y-1)\cdot r_{x,-M-1}+(y+M+1)\cdot r_{-M-1,N})\\
 &p^{H}_{x,y}=((N-x-1)\cdot r_{-M-1,y}+(x+M+1)\cdot r_{N,-M-1})\\
 &p_{x,y}=(p^{V}_{x,y}+p^{H}_{x,y}+N+M)/(2\cdot (N+M)).
 \end{split}
 \end{equation}
 When $M$ equals to $0$, the Equations (\ref{IntraGenerate1}), (\ref{IntraGenerate2}), and (\ref{IntraGenerate3})  exactly define the angular and planar prediction for the nearest reference line $L_{0}$, which is same with that defined in HEVC standard \cite{lainema2012intra}. It is noted that our prediction method with the  further reference lines is different from \cite{matsuo2009intra}. In \cite{matsuo2009intra}, the further reference line is used to replace the nearest reference line (the top row uses vertical replacement, and the left column uses horizontal replacement), and then the replaced nearest reference line is used for generating prediction. As far as the angular prediction is concerned, the method in \cite{matsuo2009intra} will not follow the assumption that the visual content follows a pure direction of propagation.

When generating the predicted block with the nearest reference line $L_{0}$, $4N+1$ pixels will be utilized. More generally, $4(N+M)+1$ pixels will be utilized for $L_{M}$ in the proposed method because of the existence of the interval. The extended pixels are  illustrated in Fig. \ref{reference_line_structure}. For each reference line, the unavailable pixel padding and the reference pixel smoothing follow the same manner with the nearest reference line in the HEVC standard \cite{HEVCdraft}. 

The reference line index $L_{M}$ will be transmitted into the bit stream.  It is coded using a similar way to code reference picture index. For example, if there are 4 reference lines,  we  use 0,10, 110, and 111 to indicate $L_{0}$ to $L_{3}$, respectively. In the proposed method, the reference line index is implemented at the CU level.  All PUs and TUs in a CU will share the same reference line index. The chroma components will reuse the downsampled reference line index of its corresponding luma component. For example, in 4:2:0 format,  $L_{0}$ is used for chroma components if $L_{0}$ or $L_{1}$ is used for luma.

\subsection{Residue Compensation}

From the structure of further reference line shown in Fig. \ref{reference_line_structure}, we can see that there exists an interval between  further reference line and the predicted block. The interval region can also be predicted by the further reference line. In addition, the  reconstructed pixels of the interval are  available if the current block is not located on the boundary of a picture. So the  residue can be estimated by subtracting its \textquotedblleft new\textquotedblright  \ prediction from its reconstruction, where the new prediction means the prediction  generated by the further reference line.  The residue of the interval is the affiliated information in utilizing  the further reference line, and can be used to calibrate the prediction because we assume that the residue also has a spatial correlation in some degree. 

\begin{figure}
	\centering
	\includegraphics[width=260pt, height=460pt]{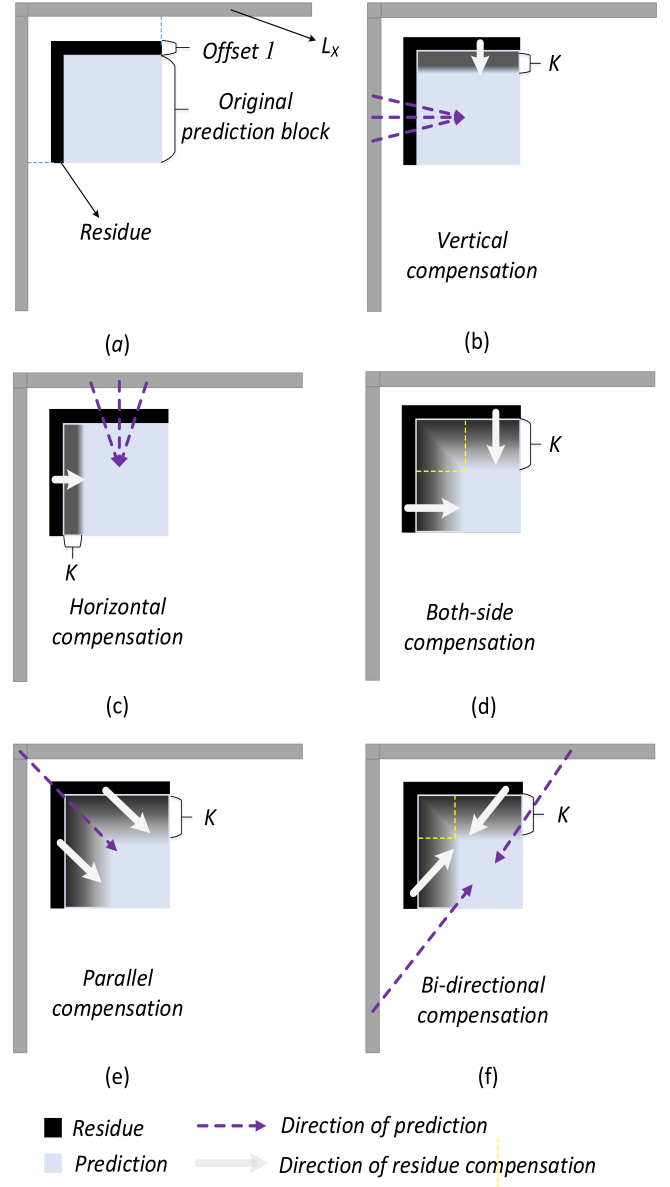}
	
	\caption{ The residue compensation types for different intra prediction modes. (a)  Residue illustration.  (b) Vertical type for mode $7\sim 13$.  (c) Horizontal type for mode $23\sim 29$. (d) Both-side type for DC and planar mode. (e) Parallel type for mode $14\sim 22$. (f) Bi-directional type for  mode $2 $ and $34$.}\label{ResidueCom}
\end{figure}

This paper proposes  compensating the  residue to the boundaries of a predicted block. Considering the assumption that spatial correlation is stronger with the distance shorter, we only compensate the residue of the nearest reference line. For this reason, we will  predict a larger block whose size equals the original block size plus one to get the nearest residue, as shown in Fig. \ref{ResidueCom} (a).
 To improve  compensation efficiency, this paper designs several  residue compensation strategies  to adapt to different intra modes, and they are described subsequently.

The vertical residue compensation is performed on the first row of a predicted block. It can be formulated as
\begin{equation}\label{ResiCom2}
p^{\bracevert}_{x,y}=	p_{x,y}+w_{y}\cdot (r_{x,-1}-p_{x,-1}), \quad\quad y=0.
\end{equation}
 $p^{\bracevert}_{x,y}$ is the predicted value after residue compensation.   $w$ indicates the compensation weight. 
In the proposed method, vertical residue compensation will be used for angular directions around horizontal intra direction (i.e. mode $7 \sim 13$ ). 
The weight $w$ is calculated  according to the  mode index:
\begin{equation}\label{weight2}
w_{y}=(14-abs(dir-HOR))\cdot 3/64, \quad\quad y=0,
\end{equation}
where the $dir$ is the current intra mode index, and $HOR$ is the horizontal mode index (i.e. $10$). 
The weight will be larger when the deviation between current intra direction and horizontal direction is smaller.
 The vertical compensation is illustrated in Fig. \ref{ResidueCom} (b).   
 
 Similarly, the horizontal residue compensation acts on the first column of a predicted block, which can be regarded as a transposition of vertical residue compensation. It is shown in  Fig. \ref{ResidueCom} (c).  
 The  horizontal residue compensation is designed for  directions around vertical intra direction (i.e. mode $23 \sim 30$ ).

The both-side residue compensation is applied on  the top and left boundaries. The top rows  are with vertical residue compensation, and the left columns are with horizontal residue compensation. It can be regarded as the combination of vertical and horizontal compensation, merely it will perform on multiple lines. It can be formulated as
\begin{equation}\label{ResiCom1}
p^{\bracevert}_{x,y}=\begin{dcases*}
	p_{x,y}+w_{y}\cdot  (r_{x,-1}-p_{x,-1}),  & $y<K$ \cr\cr
	p_{x,y}+w_{x}\cdot(r_{-1,y}-p_{-1,y}),   & $x<K$,
\end{dcases*}
\end{equation}
where $K$ is the parameter representing the number of compensated lines, and is set as 3 for both-side residue compensation in  implementation. The weight $w$ is calculated according to the distance to the residue:
\begin{equation}\label{weight1}
w_{k}=(A-k)/B, \quad \quad    k<K,
\end{equation}
where $k$ is the row or column index. $A$ and $B$ are the parameters, which are set as 3 and 4 in  implementation. 
In equation (\ref{weight1}), the compensation weight will be smaller when the pixel has a larger distance to the residue (i.e., when $k$ is larger). This is because that the residue correlation is weaker when the distance is longer. 
Fig. \ref{ResidueCom} (d) shows the illustration.
The gradually changed color indicates the intensity of  compensation weight.
In (\ref{ResiCom1}), it is noted that the top-left part ($0\leq x,y <K$, denoted by the region within dotted lines) will  perform both horizontal and vertical compensation, where the vertical compensation is prior to horizontal compensation. 
In the proposed method, the both-side  compensation  is used for  DC and planar modes.

The parallel compensation follows the same direction with the intra angular direction, as shown 
\begin{equation}\label{ResiCom3}
p^{\bracevert}_{x,y}=	p_{x,y}+w_{min(x,y)}\cdot (r_{x^{\star},y^{\star}}-p_{x^{\star},y^{\star}}),   \quad     x<K\ ||\ y<K,
\end{equation}
where the $(x^{\star},y^{\star})$ is the location by projecting the  $(x,y)$ pixel   to the nearest reference line along the angular direction. If the projection is located on a fractional position, the $p_{x^{\star},y^{\star}}$ and $r_{x^{\star},y^{\star}}$ are interpolated by the two adjacent pixels, same with the normal intra angular prediction.  The $K$ is also set to 3 for parallel compensation, and the weight is same as that in (\ref{weight1}).  The parallel compensation is illustrated in  Fig. \ref{ResidueCom} (e), and it is   performed for angular directions around diagonal direction (i.e. mode $14 \sim 22$ ).

In addition, we propose bi-directional compensation for mode $2  $ and mode $  34$, as shown in  Fig. \ref{ResidueCom} (f). The formulation is as follows
\begin{equation}\label{ResiCom4}
p^{\bracevert}_{x,y}=\begin{dcases*}
p_{x,y}+w_{y}\cdot(r_{x+y+1,-1}-p_{x+y+1,-1}), & $y<K$ \cr\cr
p_{x,y}+w_{x}\cdot(r_{-1,x+y+1}-p_{-1,x+y+1}), & $x<K$.
\end{dcases*}
\end{equation}
The $K$ is also set as 3 for bi-directional compensation in the implementation, and the weight is same as both-side compensation in (\ref{weight1}). Similarly, the top-left part ($0\leq x,y <K$) will also  perform the  compensation from both the top  and left residue, where  vertical  compensation is applied first.

\begin{figure}
	\centering
	\includegraphics[width=255pt, height=130pt]{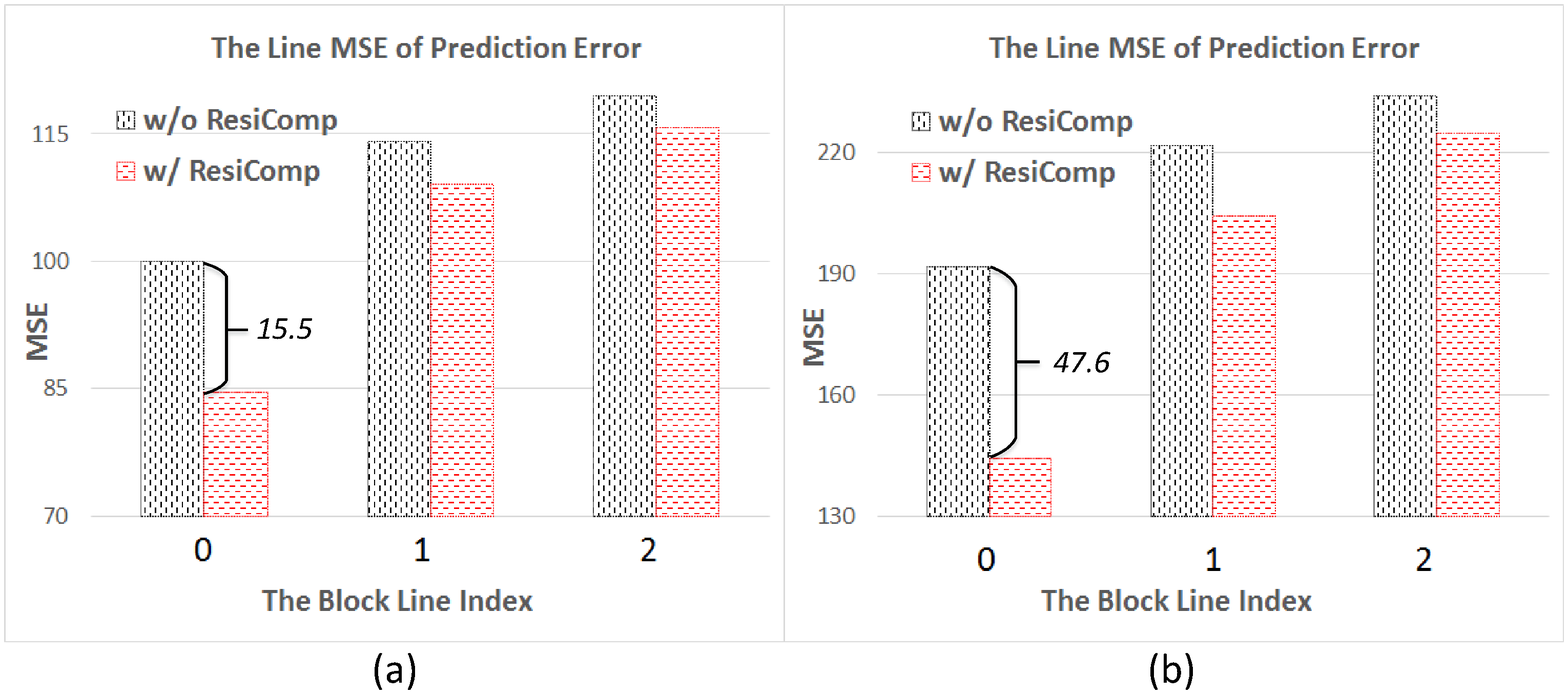}
	
	\caption{ The MSE of prediction error. The MSE is the average of each line including one row and one column. The block line index 0 indicates the most top row and most left column. The block line index 1 indicates the second top row and second left column.   The block line index 2 indicates the third top row and third left column. \textit{QP 37}. (a) $8\times8$. (b) $16\times16$. 	}\label{ResiGain}
\end{figure}
To verify the benefit of residue compensation, we calculate the  mean square error (MSE) between the  predicted block and the original block when the residue compensation is disabled or enabled. In our method, we only compensate the  three  top rows and left columns at most for all block sizes, and the compensation intensity decays from the boundary to the inner of block. Thus, we present the MSE change of the three lines, as shown in  Fig. \ref{ResiGain}. The  experiment setting of this statistic is same with that in Fig. \ref{AnaBetterRatio}.
 We merely use the compressed pixels of reference lines to predict, where the QP is 37 for compression.

For each instance in Fig. \ref{ResiGain}, the left one only uses multiples reference lines, but the residue compensation is disabled. The right one  enables the residue compensation. From the figure, we can see that the residue compensation can obviously reduce the prediction errors of the three boundary lines. The average MSE of the first block line (i.e. the most top row and most left column) can decrease 47.6 for $16\times16$ blocks if using residue compensation. It verifies that the residue can help  calibrate  prediction well on this region under the assumption that the residue also has a spatial correlation.

 After  residue compensation, we  continue to weight the prediction generated by the nearest reference line, just like the principle of bi-prediction in inter coding,  where the predictions estimated from two reference pictures are weighted. The weights are [3/4,1/4], respectively (3/4 is for the further reference line). The two predictions share the same direction. The joint utilization of further references and the nearest reference line can further refine  prediction.

\subsection{Best Reference Line Selection and Fast Algorithms}

In the encoder, the   best reference line is chosen according to the  rate distortion (RD) cost, formulated by:
\begin{equation}
L_{Best}=\arg\underset{L_{M}}{\min}\,(D_{L_{M}}+\lambda \cdot R_{L_{M}})
\end{equation}
where $D_{L_{M}}$ is the distortion between the reconstructed block using reference line $L_{M}$ and the original block, $R$ is the corresponding bits, and $\lambda$ is the  Lagrangian multiplier used in the mode decision process. The encoder will check the RD cost of using the  reference lines from the nearest one to the farthest one, and chooses the best one at the CU level. It is noted that the residue compensation is applied when checking further reference lines, and then the best reference line is chosen by rate-distortion optimization. 

In general, we can get  better prediction by using  more  reference lines. However, the encoding complexity will  almost linearly increase with the number of checked reference lines.
Considering the tradeoff between  compression efficiency and encoding complexity, we  use at most 4 reference lines (i.e. from $L_{0}$ to $L_{3}$) in the proposed method. The additional three reference lines are required to be cached to support the proposed method. However, because of the existence of deblocking procedure, these additional pixels have already been cached \cite{norkin2012hevc}. Thus, no additional memory increase is introduced by the proposed method.

Sometimes, the complexity for checking 4 reference lines still may  not be welcome in some situations.
For this reason,   this paper also provides a fast solution which is incorporated with  several acceleration algorithms. These   algorithms are subsequently introduced.

\subsubsection{Subset Selection}
\begin{figure}
	\centering
	\includegraphics[width=250pt, height=160pt]{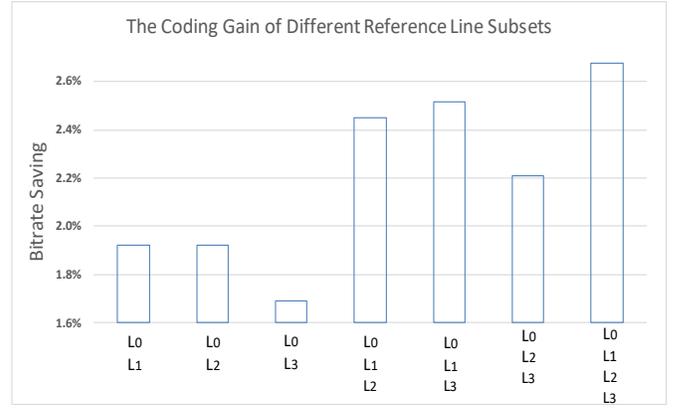}
	
	\caption{The coding gain of using different reference line subsets. One frame. }\label{PickSubset}
\end{figure}

Considering the tradeoff between  compression efficiency and encoding complexity, we propose  using the near 4 reference lines, namely from $L_{0}$ to $L_{3}$, in the full search scheme. However, we propose  selecting a subset of the 4 lines to accelerate the encoder. As for the selection of the subset, we make the choice according to  experimental results. When considering that the nearest reference line $L_{0}$ has the strongest statistical correlation, we keep it in the subset initially. Then we test all  candidates of the subset. They are \{$L_{0}$, $L_{1}$\}, \{$L_{0}$, $L_{2}$\}, \{$L_{0}$, $L_{3}$\}, \{$L_{0}$, $L_{1}$, $L_{2}$\}, \{$L_{0}$, $L_{1}$, $L_{3}$\}, and \{$L_{0}$, $L_{2}$,  $L_{3}$\}. For convenience, we only test one frame of five sequences (\textit{BasketballDrill, BQTerrace, Cactus, Traffic,}  and \textit{PeopleOnStreet}). The experimental results are shown in Fig. \ref{PickSubset}. 
As to the coding efficiency, we can see that the result of subset \{$L_{0}$, $L_{1}$, $L_{3}$\} performs best when using three reference lines, and it has $0.59\%$ gain over that of \{$L_{0}$, $L_{1}$\}, which performs best when using two reference lines. Moreover, the result of subset \{$L_{0}$, $L_{1}$, $L_{3}$\} only has a $0.16\%$ loss over that of full set \{$L_{0}$, $L_{1}$, $L_{2}$, $L_{3}$\}. Meanwhile, the relation between encoding complexity and the number of checked reference lines is almost linear.
From this observation, we propose  using the subset \{$L_{0}$, $L_{1}$, $L_{3}$\}  in the fast solution, namely skipping the $L_{2}$. It should be noted that the disabling $L_{2}$ in the fast solution is a normative modification (i.e. there needs corresponding modification in the decoder) because we need to modify the binarization of reference line index.

\subsubsection{Block Size Decision}
In the hierarchical block structure of HEVC, the size of CU varies from $8\times8$ to  $64\times64$. However, for intra coding, the larger  blocks (i.e. 64x64 and 32x32) are  chosen less often because   the larger  blocks  are most likely used in the regions whose textures are very simple (e.g. smooth regions). In the proposed method, to accelerate the encoder, the  CU size of 64x64 is  not checked for further reference lines.

 For regions full of complex textures, the encoder generally prefers the smaller CU for more elaborate prediction. To know whether the current CU locates a  region with complex textures, we use the sizes of neighbor blocks as a hint.
 If the neighbor blocks are  small in size, it means that the current region is likely to be full of texture, and the current CU probably also chooses small size. For this reason, we will not check the further reference lines for $32\times 32$ CU in the fast solution when the sizes of the above and left PUs are both less than 16. 
\begin{figure}
	\centering
	\includegraphics[width=260pt, height=420pt]{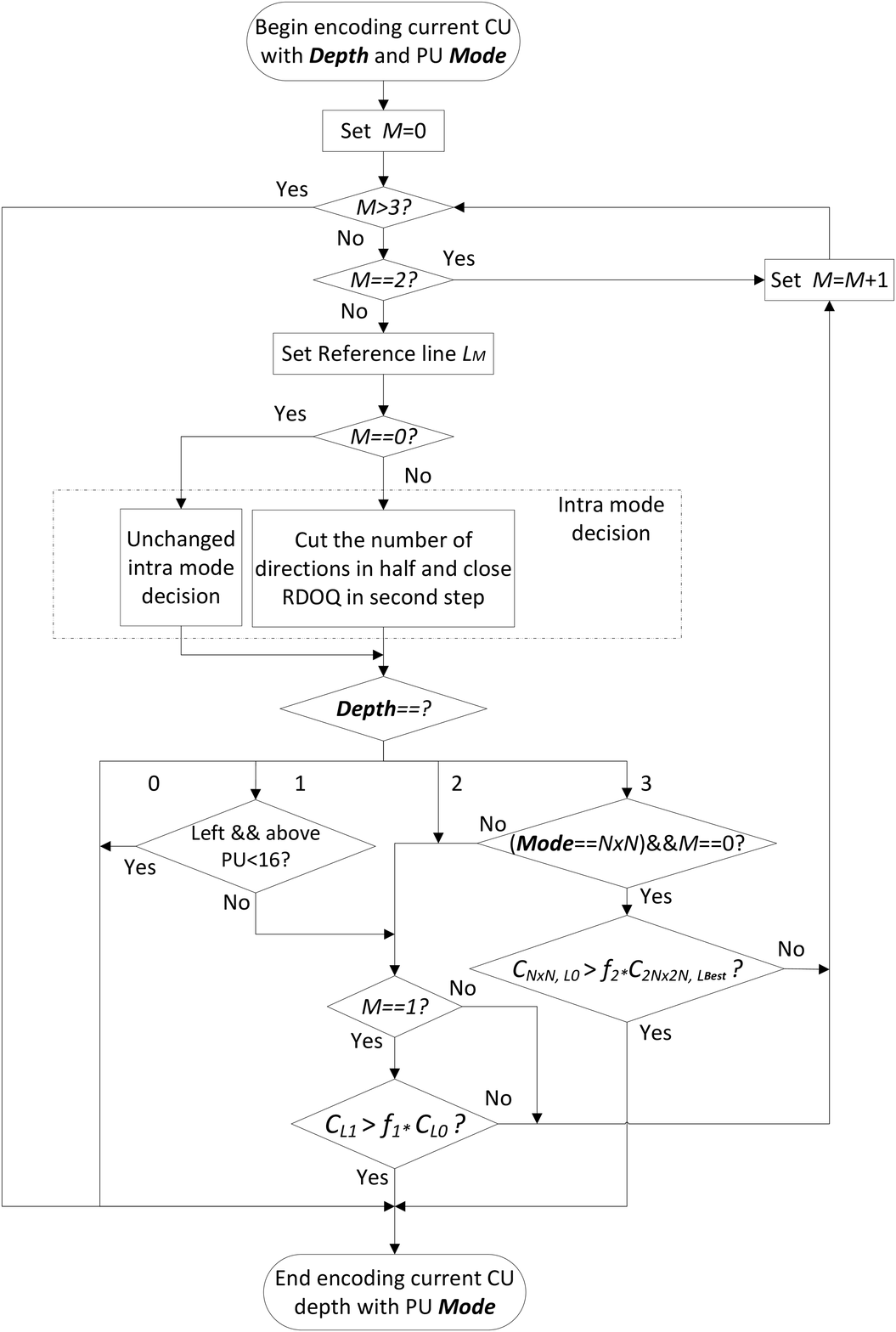}
	\caption{ The flowchart of the proposed fast algorithms.}\label{flowcharFast}
\end{figure}
\subsubsection{RD Cost-Based Decision}
Because of the strong spatial correlation, sometimes the nearest reference line $L_{0}$ already can  predict the original block well. In this situation, the check of further reference lines seems to be superfluous. Thus, we propose  skipping the check of further reference lines when the nearest reference line is  efficient enough, and we use the RD costs of $L_{0}$ and $L_{1}$ as the hint. We will not check the reference lines which are further than  $L_{1}$ if we find that the RD cost of  $L_{1}$ is  larger than that of  $L_{0}$ to some degree. This condition can be expressed by
\begin{equation}\label{CSkip}
C_{L_{1}}>f_{1}\times C_{L_{0}},
\end{equation}
where $C_{L_{M}}$ indicates the rate distortion cost of CU with reference line $L_{M}$ and the $f_{1}$ is a parameter, which is set as $1.1$ in the implementation. 

In the encoder of HEVC reference software (HM), the intra mode  $2N\times2N$ will be checked before the mode $N\times N$.  In this procedure, we can skip  checking the further reference lines of  mode $N\times N$ if we find that the mode $2N\times2N$ probably performs better than the mode $N\times N$. We use the RD cost of mode  $N\times N$ with the nearest reference line $L_{0}$  and  that of mode $2N\times2N$  with the best reference line $L_{Best}$ as the hint. If  the following condition is satisfied, the further reference lines of mode $N\times N$ will not be checked because we deduce that the mode $2N\times2N$ performs better than mode $N\times N$.
\begin{equation}\label{CSkip2}
C_{N\times N \verb|,| L_{0}}>f_{2} \times C_{2N\times 2N \verb|,|  L_{Best}}.
\end{equation}
 Here, $f_{2}$ is set as $1.2$ in the implementation.

\subsubsection{Other Accelerations}
The whole procedure of intra mode decision in the encoder of HM can be divided into three main steps \cite{zhao2011fast}:  rough mode decision (RMD) based on SATD  cost, fine-grained direction decision based on RD cost with fixed TU size, and the check of best residual quadtree (RQT) size with the best direction. The second step will occupy  a large percentage  of the whole encoding complexity as several full encoding procedures (including prediction, transform, quantization and entropy encoding) will be performed. To accelerate the proposed scheme, the number of direction candidates  will be  decreased by half in the second step for the further reference lines. In addition, the tool of rate distortion optimized quantization (RDOQ) is disabled in the second step for further acceleration. 

It should be noted that the all of the proposed fast algorithms  only work for  further reference lines. The intra coding with the nearest reference line $L_{0}$ remains unchanged. The flowchart of the proposed fast algorithms is shown in Fig. \ref{flowcharFast}.

\section{Experimental Results}
%single 1.2, double 1.34
\begin{table*}\renewcommand{\arraystretch}{1.34}
	\centering
	\caption{Bitrate Saving of The Proposed Scheme}
	\label{main_result}
	\begin{tabular*}{500pt}{@{\extracolsep{\fill}}cccccccccccccc}	 
		\toprule[1.1pt]
		%the above are for single column
		%\multirow{2}{*}{Sequence}  & \multicolumn{3}{c}{Full Search-Lossy} & & \multicolumn{3}{c}{Fast Search-Lossy} &\multicolumn{1}{c}{Full Search}&\multicolumn{1}{c}{Fast Search}\\
		%& Y & Cb  & Cr & &  Y & Cb  & Cr&-Lossless	&-Lossless 	\\
		\multirow{2}{*}{Sequence}  & \multicolumn{3}{c}{Full Search-Lossy} & & \multicolumn{3}{c}{Fast Search-Lossy} &\multirow{2}{*}{Full Search-Lossless}&\multirow{2}{*}{Fast Search-Lossless}\\
		& Y & Cb  & Cr & &  Y & Cb  & Cr	 	\\
		\hline
		
		\textit{Traffic}&		              $ -2.4\%	$  &       $ -1.6\%	$   &       $ -1.8\%	$     &     &    $ -2.1\%	$  &       $ -1.1\%	$   &       $ -1.2\%	$   &       $ -1.0\%	$   &       $ -0.9\%	$    \\
		\textit{PeopleOnStreet}&	          $ -2.6\%	$  &       $ -2.5\%	$   &       $ -2.5\%	$     &     &    $ -2.3\%	$  &       $ -1.8\%	$   &       $ -1.8\%	$   &       $ -1.4\%	$   &       $ -1.2\%	$    \\
		\textit{Nebuta}&	                  $ -2.3\%	$  &       $ -2.5\%	$   &       $ -1.9\%	$     &     &    $ -1.9\%	$  &       $ -2.0\%	$   &       $ -1.5\%	$   &       $ -2.1\%	$   &       $ -1.9\%	$    \\
		\textit{SteamLocomotive}&             $ -1.4\%	$  &       $ -2.6\%	$   &       $ -2.9\%	$     &     &    $ -1.1\%	$  &       $ -0.8\%	$   &       $ -1.0\%	$   &       $ -2.1\%	$   &       $ -1.8\%	$    \\
		\textit{Kimono}&	                  $ -1.6\%	$  &       $ -1.4\%	$   &       $ -1.5\%	$     &     &    $ -1.4\%	$  &       $ -0.9\%	$   &       $ -0.9\%	$   &       $ -1.0\%	$   &       $ -0.9\%	$    \\
		\textit{ParkScene}&                   $ -2.0\%	$  &       $ -1.2\%	$   &       $ -1.5\%	$     &     &    $ -1.7\%	$  &       $ -0.7\%	$   &       $ -1.0\%	$   &       $ -1.2\%	$   &       $ -1.0\%	$    \\
		\textit{Cactus}&				      $ -2.5\%	$  &       $ -1.4\%	$   &       $ -2.2\%	$     &     &    $ -2.1\%	$  &       $ -0.8\%	$   &       $ -1.3\%	$   &       $ -1.2\%	$   &       $ -1.0\%	$    \\
		\textit{BasketballDrive}&             $ -2.6\%	$  &       $ -2.4\%	$   &       $ -2.5\%	$     &     &    $ -2.2\%	$  &       $ -1.2\%	$   &       $ -1.4\%	$   &       $ -1.4\%	$   &       $ -1.1\%	$    \\
		\textit{BQTerrace}&		              $ -2.8\%	$  &       $ -2.2\%	$   &       $ -2.5\%	$     &     &    $ -2.2\%	$  &       $ -1.3\%	$   &       $ -1.5\%	$   &       $ -1.3\%	$   &       $ -1.1\%	$    \\
		\textit{BasketballDrill}&             $ -3.7\%	$  &       $ -4.0\%	$   &       $ -4.3\%	$     &     &    $ -3.2\%	$  &       $ -2.9\%	$   &       $ -3.0\%	$   &       $ -1.1\%	$   &       $ -1.0\%	$    \\
		\textit{BQMall}&                      $ -2.0\%	$  &       $ -1.4\%	$   &       $ -1.6\%	$     &     &    $ -1.7\%	$  &       $ -0.9\%	$   &       $ -1.0\%	$   &       $ -1.1\%	$   &       $ -0.9\%	$    \\
		\textit{PartyScene}&                  $ -2.0\%	$  &       $ -1.2\%	$   &       $ -1.3\%	$     &     &    $ -1.5\%	$  &       $ -0.7\%	$   &       $ -0.8\%	$   &       $ -1.1\%	$   &       $ -0.9\%	$    \\
		\textit{RaceHorsesC}&	              $ -2.2\%	$  &       $ -1.7\%	$   &       $ -2.1\%	$     &     &    $ -1.8\%	$  &       $ -1.2\%	$   &       $ -1.5\%	$   &       $ -1.1\%	$   &       $ -1.0\%	$    \\
		\textit{BasketballPass}&	          $ -2.2\%	$  &       $ -1.9\%	$   &       $ -2.0\%	$     &     &    $ -1.9\%	$  &       $ -1.1\%	$   &       $ -1.2\%	$   &       $ -1.1\%	$   &       $ -0.9\%	$    \\
		\textit{BQSquare}&	                  $ -1.9\%	$  &       $ -1.0\%	$   &       $ -1.2\%	$     &     &    $ -1.4\%	$  &       $ -0.6\%	$   &       $ -0.8\%	$   &       $ -1.1\%	$   &       $ -0.9\%	$    \\
		\textit{BlowingBubbles}&	          $ -1.9\%	$  &       $ -1.2\%	$   &       $ -1.3\%	$     &     &    $ -1.5\%	$  &       $ -0.7\%	$   &       $ -0.8\%	$   &       $ -0.9\%	$   &       $ -0.8\%	$    \\
		\textit{RaceHorses}&	              $ -2.1\%	$  &       $ -2.1\%	$   &       $ -2.1\%	$     &     &    $ -1.9\%	$  &       $ -1.4\%	$   &       $ -1.5\%	$   &       $ -1.1\%	$   &       $ -0.9\%	$    \\
		\textit{FourPeople}&	              $ -2.1\%	$  &       $ -1.6\%	$   &       $ -1.7\%	$     &     &    $ -1.9\%	$  &       $ -1.1\%	$   &       $ -1.1\%	$   &       $ -1.2\%	$   &       $ -1.0\%	$    \\
		\textit{Johnny}&	                  $ -2.3\%	$  &       $ -3.0\%	$   &       $ -2.7\%	$     &     &    $ -2.0\%	$  &       $ -1.9\%	$   &       $ -1.8\%	$   &       $ -1.2\%	$   &       $ -1.0\%	$    \\
		\textit{KristenAndSara}&	          $ -2.1\%	$  &       $ -2.6\%	$   &       $ -2.4\%	$     &     &    $ -1.8\%	$  &       $ -1.7\%	$   &       $ -1.4\%	$   &       $ -1.2\%	$   &       $ -1.1\%	$    \\
		\textit{BaskeballDrillText}&	      $ -3.2\%	$  &       $ -3.2\%	$   &       $ -3.3\%	$     &     &    $ -2.8\%	$  &       $ -2.3\%	$   &       $ -2.3\%	$   &       $ -1.1\%	$   &       $ -0.9\%	$    \\
		\textit{ChinaSpeed}&	              $ -1.8\%	$  &       $ -1.1\%	$   &       $ -1.2\%	$     &     &    $ -1.4\%	$  &       $ -0.6\%	$   &       $ -0.7\%	$   &       $ -0.7\%	$   &       $ -0.6\%	$    \\
		\textit{SlideEditing}&	              $ -2.2\%	$  &       $ -1.2\%	$   &       $ -1.2\%	$     &     &    $ -1.6\%	$  &       $ -0.8\%	$   &       $ -0.7\%	$   &       $ -1.0\%	$   &       $ -0.7\%	$    \\
		\textit{SlideShow}&	                  $ -1.8\%	$  &       $ -1.1\%	$   &       $ -1.1\%	$     &     &    $ -1.5\%	$  &       $ -0.8\%	$   &       $ -0.8\%	$   &       $ -1.3\%	$   &       $ -1.1\%	$    \\
		
		\textit{\textbf{Average-CTC}}&        $  -\textbf{2.2\%}  $        &          $  -\textbf{1.9\%}    $   &          $  -\textbf{2.0\%}    $       &      &    $  -\textbf{1.9\%}    $    &          $  -\textbf{1.2\%}    $    &          $  -\textbf{1.3\%}    $    &          $  -\textbf{1.2\%}    $  &          $  -\textbf{1.0\%}    $      \\

		\textit{Tango}&	                      $ -2.5\%	$  &       $ -3.3\%	$   &       $ -2.7\%	$     &     &   $ -1.9\%	$  &       $ -1.8\%	$   &       $ -1.3\%	$   &       $ -1.1\%	$   &       $ -1.0\%	$    \\
		\textit{Drums100}&	                  $ -2.4\%	$  &       $ -3.1\%	$   &       $ -3.1\%	$     &     &   $ -2.1\%	$  &       $ -2.2\%	$   &       $ -2.2\%	$   &       $ -1.4\%	$   &       $ -1.3\%	$    \\
		\textit{CampfireParty}&	 	          $ -2.7\%	$  &       $ -2.0\%	$   &       $ -2.3\%	$     &     &   $ -2.3\%	$  &       $ -1.1\%	$   &       $ -1.3\%	$   &       $ -1.1\%	$   &       $ -1.0\%	$    \\
		\textit{ToddlerFountain}&             $ -1.7\%	$  &       $ -1.1\%	$   &       $ -1.1\%	$     &     &   $ -1.5\%	$  &       $ -0.4\%	$   &       $ -0.6\%	$   &       $ -1.0\%	$   &       $ -0.9\%	$    \\
		\textit{CatRobot}&	                  $ -2.6\%	$  &       $ -2.3\%	$   &       $ -2.5\%	$     &     &   $ -2.3\%	$  &       $ -1.4\%	$   &       $ -1.6\%	$   &       $ -1.0\%	$   &       $ -0.8\%	$    \\
		\textit{TrafficFlow}&	              $ -4.3\%	$  &       $ -4.5\%	$   &       $ -4.6\%	$     &     &   $ -3.7\%	$  &       $ -3.1\%	$   &       $ -3.0\%	$   &       $ -0.9\%	$   &       $ -0.8\%	$    \\
		\textit{DaylightRoad}&	              $ -3.1\%	$  &       $ -2.2\%	$   &       $ -2.0\%	$     &     &   $ -2.7\%	$  &       $ -1.2\%	$   &       $ -0.8\%	$   &       $ -1.0\%	$   &       $ -0.8\%	$    \\
		\textit{Rollercoaster}&	 	          $ -2.2\%	$  &       $ -2.7\%	$   &       $ -2.3\%	$     &     &   $ -1.7\%	$  &       $ -1.8\%	$   &       $ -1.5\%	$   &       $ -1.0\%	$   &       $ -0.9\%	$    \\

		\textit{\textbf{Average-4K}}&        $  -\textbf{2.7\%}  $        &          $  -\textbf{2.7\%}    $   &          $  -\textbf{2.6\%}    $    &    &      $  -\textbf{2.3\%}    $    &          $  -\textbf{1.6\%}    $    &          $  -\textbf{1.5\%}    $    &          $  -\textbf{1.1\%}    $    &          $  -\textbf{0.9\%}    $       \\
		\textit{\textbf{Average-All}}&       $  -\textbf{2.3\%}  $        &          $  -\textbf{2.1\%}    $   &          $  -\textbf{2.2\%}    $    &    &      $  -\textbf{2.0\%}    $    &          $  -\textbf{1.3 \%}  $    &          $  -\textbf{1.4\%}    $    &          $  -\textbf{1.2\%}    $    &          $  -\textbf{1.0\%}    $       \\

		\textit{\textbf{Encoding / Decoding  time}}&     \multicolumn{3}{c}{$  \textbf{463\%/112\%}  $ }        &   &     \multicolumn{3}{c}{$  \textbf{212\%/110\%}   $ }        &          $  \textbf{477\%/113\%}    $    &          $  \textbf{275\%/110\%}    $      \\

		\bottomrule	[1.1pt]	
	\end{tabular*}
\end{table*}
\subsection{Experiment Setting}

To verify the compression performance of the proposed multiple line-based intra prediction scheme, we implement it into the  HEVC reference software HM-16.9 \cite{hm}. The test sequences include the whole range of HEVC standard test  sequences in  common test conditions (CTC) \cite{ctc}, which are  specified by JCT-VC (Joint Collaborative Team on Video Coding). They are arranged into five classes: Class A (2560$\times$1600), Class B (1080p), Class C (WVGA), Class D (WQVGA), Class E (720p), and Class F (screen content). In addition, we choose  eight 4K sequences from  \cite{JVETctc} to verify  the proposed method on higher resolution sequences. The first 32 frames are tested in the eight 4K sequences.

As our algorithms are designed for intra coding, we only test the  all intra main configuration. The quantization parameters are set as 22, 27, 32, and 37. The   parameter settings of other tools    strictly follow the HEVC CTC, unless  pointed out  explicitly. The results are evaluated by BD-Rate \cite{psnr}, where a negative number indicates bitrate saving and a positive number indicates an increasing bitrate. Meanwhile, we also test our algorithms under lossless coding,  where  coding efficiency is measured by bitrate change.

\subsection{Experimental Results of The Proposed Method}

The experimental results of the proposed multiple line-based scheme with full search (i.e. disabling the proposed fast algorithms)  are shown in Table I. The average bitrate saving of all sequences is about $2.3\%$, and the  maximum bitrate saving is  $4.3\%$ for \textit{TrafficFlow}. For the 4K sequences, the proposed method can achieve an average gain of $2.7\%$. As for computational complexity, we can see that the
proposed  scheme with full search will increase encoding time by about $363\%$. This is because the encoder needs to  check the intra coding with 4 reference lines. In addition, the post-processing of residue compensation needs extra interpolations. Similarly, as the existence of residue compensation, the decoding time has a small increase.

In addition, we conduct the experimental results under lossless coding, also shown in Table I. From the results, we can see that the proposed full search method achieves an average bitrate saving of $1.2\%$. When compared with the $2.3\%$ of lossy coding, it is smaller because  the advantage that the further reference lines have  relatively better quality does not exist  under lossless coding. But due to the incoherence caused by signal noise or the texture of other object, the utilization of multiple lines still performs better than only using single line, and achieves notable $1.2\%$ gain. In particular, the \textit{Nebuta} achieve the largest bitrate saving, up to $2.1\%$. We look into this  sequence, and find that it is full of noise. This make the advantage of multiple line-based scheme more prominent.

In this paper, we also provide an alternative solution that cooperates with several fast algorithms. The experimental results of the proposed scheme with fast  algorithms are also shown in Table \ref{main_result}. From the table, we can see that the increased encoding time has an obvious reduction, from $363\%$  to $112\%$. Nevertheless,  coding efficiency only has a small drop. The average coding gain of lossy coding  still reaches $2.0\%$,  where the maximum bitrate saving is  $3.7\%$ for \textit{TrafficFlow}. For lossless coding, there is only a $0.2\%$ loss when compared to the full search scheme. These results verify that the proposed fast algorithms  are effective.

\subsection{More Analyses}

\begin{figure}
	\centering
	\includegraphics[width=250pt, height=165pt]{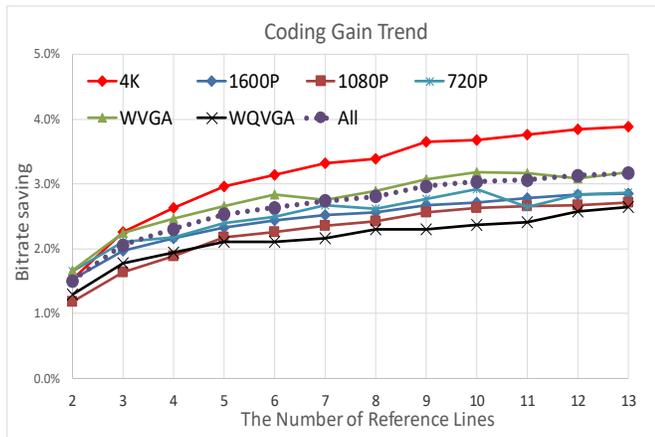}	
	\caption{The trend of coding gain when the  number of reference lines is different, one frame.}\label{CodingGainTrendEachClass}
\end{figure}

Considering the tradeoff between coding efficiency and encoding complexity, this paper uses 4 reference lines at most.
From the results shown in Table I, we can see that the 4 line-based intra prediction can significantly improve  coding efficiency.
For more analyses, we collect the results when the number of reference lines is different. 
In particular, the performances of sequences with different spatial resolutions are grouped separately to investigate the relationship between the  appropriate  number of reference lines and the resolution. The detailed results are presented in Fig. \ref{CodingGainTrendEachClass}.
 From the figure, we can see that the performance of sequences with all resolutions has relatively large improvement when the number of reference lines is in the early stage of growth (e.g. from $2$ to $4$), even for sequences with lower resolution (e.g., WQVGA). For this reason, in the proposed method, the maximum  number of reference lines is fixed as 4 for all sequences with different resolutions. From the average results of all sequences, we can see that the performance can continue to increase even the number goes to $13$. The bit saving of checking $13$ lines is $3.2\%$ on average. However, when the number of reference lines go larger, the performance increases more slowly, and there even exists little fluctuation in some resolutions. This is because that the overhead of transmitting the reference line index will be heavier, which constrains the increase of performance. 

\begin{figure}
	\centering
	\includegraphics[width=250pt, height=168pt]{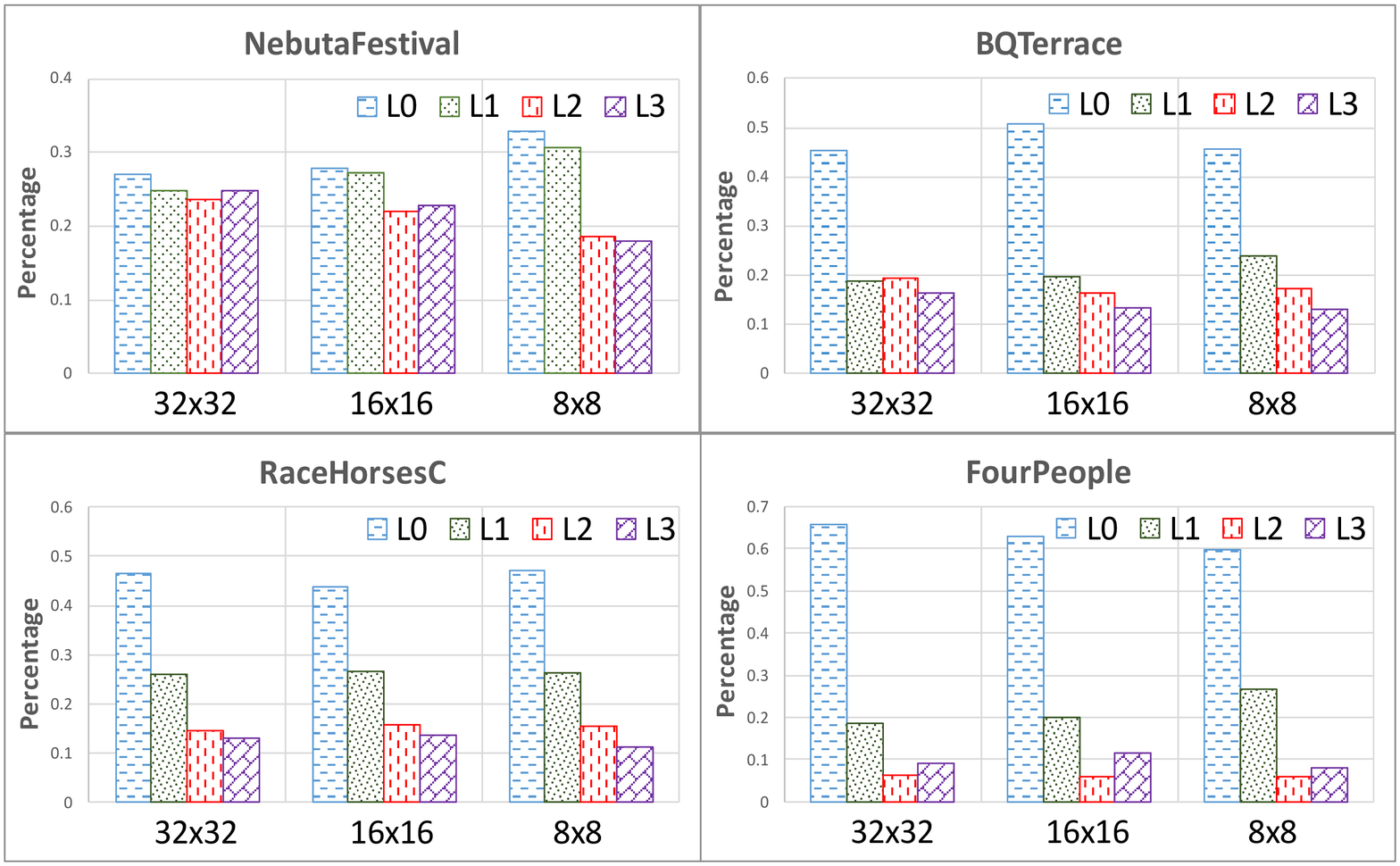}
	
	\caption{ The percentage distribution of reference lines, full search. QP: $27$.  (a) \textit{NebutaFestival.} (b)  \textit{BQTerrace.} (c) \textit{RaceHorsesC.} (d) \textit{FourPeople}.}\label{RefLineRatio}
\end{figure}
To verify the effectiveness of the proposed method, we also  collect the  reference line distributions from the actual bit streams, different with the analysis in Fig. \ref{AnaBetterRatio}, which is conducted outside the codec framework.  Fig. \ref{RefLineRatio} shows the results of four sequences from different classes. From the figure, we can find that the percentage follows similar distribution for different CU sizes from $32\times 32$ to $8\times8$,  where we do not collect the information of $64\times 64$ blocks as they are seldom chosen. For \textit{NebutaFestival}, there are $72.1\%$  $16\times16$ blocks choosing  further reference lines. The  large percentage verifies the effectiveness of the proposed method.  In all of these examples, the $L_{0}$ is still the most chosen  because $L_{0}$ has the strongest spatial correlation.

\section{Conclusions}
This paper proposes a multiple line-based intra prediction scheme  
to improve  video coding efficiency by utilizing  local further reference lines. A residue compensation procedure is introduced to calibrate  prediction along the block boundaries when using  further reference lines. 
Experimental results show that the proposed full search algorithm improves coding efficiency by $2.3\%$ on average and up to $4.3\%$. In addition, this paper designs several acceleration algorithms. When enabling these fast algorithms, the encoding time only increases about $112\%$, but the average bit saving  still reaches  $2.0\%$, where the maximum is $3.7\%$.

{
%\small	
\footnotesize 
\bibliographystyle{IEEEtran}
\bibliography{paper}

\begin{thebibliography}{10}
\providecommand{\url}[1]{#1}
\csname url@samestyle\endcsname
\providecommand{\newblock}{\relax}
\providecommand{\bibinfo}[2]{#2}
\providecommand{\BIBentrySTDinterwordspacing}{\spaceskip=0pt\relax}
\providecommand{\BIBentryALTinterwordstretchfactor}{4}
\providecommand{\BIBentryALTinterwordspacing}{\spaceskip=\fontdimen2\font plus
\BIBentryALTinterwordstretchfactor\fontdimen3\font minus
  \fontdimen4\font\relax}
\providecommand{\BIBforeignlanguage}[2]{{%
\expandafter\ifx\csname l@#1\endcsname\relax
\typeout{** WARNING: IEEEtran.bst: No hyphenation pattern has been}%
\typeout{** loaded for the language `#1'. Using the pattern for}%
\typeout{** the default language instead.}%
\else
\language=\csname l@#1\endcsname
\fi
#2}}
\providecommand{\BIBdecl}{\relax}
\BIBdecl

\bibitem{sullivan2012overview}
G.~J. Sullivan, J.~Ohm, W.-J. Han, and T.~Wiegand, ``{Overview of the high
  efficiency video coding (HEVC) standard},'' \emph{IEEE Transactions on
  Circuits and Systems for Video Technology}, vol.~22, no.~12, pp. 1649--1668,
  2012.

\bibitem{ohm2012comparison}
J.-R. Ohm, G.~J. Sullivan, H.~Schwarz, T.~K. Tan, and T.~Wiegand, ``{Comparison
  of the coding efficiency of video coding standards —including high
  efficiency video coding (HEVC)},'' \emph{IEEE Transactions on Circuits and
  Systems for Video Technology}, vol.~22, no.~12, pp. 1669--1684, 2012.

\bibitem{lainema2012intra}
J.~Lainema, F.~Bossen, W.-J. Han, J.~Min, and K.~Ugur, ``{Intra coding of the
  HEVC standard},'' \emph{IEEE Transactions on Circuits and Systems for Video
  Technology}, vol.~22, no.~12, pp. 1792--1801, 2012.

\bibitem{chen2015new}
C.~Chen, B.~Zeng, S.~Zhu, Z.~Miao, and L.~Zeng, ``{A new block-based coding
  method for HEVC intra coding},'' in \emph{2015 IEEE International Conference
  on Multimedia \& Expo Workshops (ICMEW)}.\hskip 1em plus 0.5em minus
  0.4em\relax IEEE, 2015, pp. 1--6.

\bibitem{HEVCdraft}
G.~J. S. J.-R.~O. B.~Bross, W.-J.~Han and T.~Wiegand, \emph{{High Efficiency
  Video Coding (HEVC) Text Specification Draft 7}}, JCTVC-I1003, 9th Meeting:
  Geneva, May 2012.

\bibitem{kamisli2012intra}
F.~Kamisli, ``Intra prediction based on statistical modeling of images,'' in
  \emph{Visual Communications and Image Processing (VCIP), 2012 IEEE}.\hskip
  1em plus 0.5em minus 0.4em\relax IEEE, 2012, pp. 1--6.

\bibitem{kamisli2013intra}
F.~\vspace{0mm} Kamisli, ``Intra prediction based on markov process modeling of
  images,'' \emph{IEEE Transactions on Image Processing}, vol.~22, no.~10, pp.
  3916--3925, 2013.

\bibitem{kamisli2015block}
F.~Kamisli, ``{Block-based spatial prediction and transforms based on 2D Markov
  processes for image and video compression},'' \emph{IEEE Transactions on
  Image Processing}, vol.~24, no.~4, pp. 1247--1260, 2015.

\bibitem{chen2013recursive}
Y.~Chen, J.~Han, and K.~Rose, ``A recursive extrapolation approach to intra
  prediction in video coding,'' in \emph{2013 IEEE International Conference on
  Acoustics, Speech and Signal Processing (ICASSP)}.\hskip 1em plus 0.5em minus
  0.4em\relax IEEE, 2013, pp. 1734--1738.

\bibitem{li2014rate}
S.~Li, Y.~Chen, J.~Han, T.~Nanjundaswamy, and K.~Rose, ``Rate-distortion
  optimization and adaptation of intra prediction filter parameters,'' in
  \emph{2014 IEEE International Conference on Image Processing (ICIP)}.\hskip
  1em plus 0.5em minus 0.4em\relax IEEE, 2014, pp. 3146--3150.

\bibitem{liu2007image}
D.~Liu, X.~Sun, F.~Wu, S.~Li, and Y.-Q. Zhang, ``Image compression with
  edge-based inpainting,'' \emph{IEEE Transactions on Circuits and Systems for
  Video Technology}, vol.~17, no.~10, p. 1273, 2007.

\bibitem{liu2008edge}
D.~Liu, X.~Sun, F.~Wu, and Y.-Q. Zhang, ``Edge-oriented uniform intra
  prediction,'' \emph{IEEE Transactions on Image Processing}, vol.~17, no.~10,
  pp. 1827--1836, 2008.

\bibitem{doshkov2010towards}
D.~Doshkov, P.~Ndjiki-Nya, H.~Lakshman, M.~K{\"o}ppel, and T.~Wiegand,
  ``Towards efficient intra prediction based on image inpainting methods,'' in
  \emph{Picture Coding Symposium (PCS), 2010}.\hskip 1em plus 0.5em minus
  0.4em\relax IEEE, 2010, pp. 470--473.

\bibitem{zhang2014improving}
Y.~Zhang and Y.~Lin, ``{Improving HEVC intra prediction with PDE-based
  inpainting},'' in \emph{Asia-Pacific Signal and Information Processing
  Association, 2014 Annual Summit and Conference (APSIPA)}.\hskip 1em plus
  0.5em minus 0.4em\relax IEEE, 2014, pp. 1--5.

\bibitem{qi2012intra}
X.~Qi, T.~Zhang, F.~Ye, A.~Men, and B.~Yang, ``{Intra prediction with enhanced
  inpainting method and vector predictor for HEVC},'' in \emph{2012 IEEE
  International Conference on Acoustics, Speech and Signal Processing
  (ICASSP)}.\hskip 1em plus 0.5em minus 0.4em\relax IEEE, 2012, pp. 1217--1220.

\bibitem{lai2015error}
Y.-H. Lai and Y.~Lin, ``{Error diffused intra prediction for HEVC},'' in
  \emph{2015 IEEE International Conference on Acoustics, Speech and Signal
  Processing (ICASSP)}.\hskip 1em plus 0.5em minus 0.4em\relax IEEE, 2015, pp.
  1424--1427.

\bibitem{zhang2011novel}
L.~Zhang, X.~Zhao, S.~Ma, Q.~Wang, and W.~Gao, ``Novel intra prediction via
  position-dependent filtering,'' \emph{Journal of Visual Communication and
  Image Representation}, vol.~22, no.~8, pp. 687--696, 2011.

\bibitem{chen2016improving}
H.~Chen, T.~Zhang, M.-T. Sun, A.~Saxena, and M.~Budagavi, ``Improving intra
  prediction in high efficiency video coding,'' \emph{IEEE Transactions on
  Image Processing}, vol.~25, no.~8, pp. 3671--3682, 2016.

\bibitem{BiIntra}
A.~T. T.~Shiodera and T.~Chujoh, \emph{“Bidirectional Intra Prediction”},
  ITU-T, Marrakech, Morocco, Tech. Rep. VCEG-AE14, Jan. 2007.

\bibitem{ye2008improved}
Y.~Ye and M.~Karczewicz, ``Improved {H.264} intra coding based on
  bi-directional intra prediction, directional transform, and adaptive
  coefficient scanning,'' in \emph{2008 15th IEEE International Conference on
  Image Processing}.\hskip 1em plus 0.5em minus 0.4em\relax IEEE, 2008, pp.
  2116--2119.

\bibitem{yeh2015predictive}
C.-H. Yeh, T.-Y. Tseng, C.-W. Lee, and C.-Y. Lin, ``Predictive texture
  synthesis-based intra coding scheme for advanced video coding,'' \emph{IEEE
  Transactions on Multimedia}, vol.~17, no.~9, pp. 1508--1514, 2015.

\bibitem{tan2006intra}
T.~K. Tan, C.~S. Boon, and Y.~Suzuki, ``Intra prediction by template
  matching,'' in \emph{Image Processing, 2006 IEEE International Conference
  on}.\hskip 1em plus 0.5em minus 0.4em\relax IEEE, 2006, pp. 1693--1696.

\bibitem{tan2007intra}
T.~K. \vspace{0mm} Tan, C.~S. Boon, and Y.~Suzuki, ``Intra prediction by
  averaged template matching predictors,'' in \emph{2007 4th IEEE Consumer
  Communications and Networking Conference}.\hskip 1em plus 0.5em minus
  0.4em\relax IEEE, 2007, pp. 405--409.

\bibitem{guo2008priority}
Y.~Guo, Y.-K. Wang, and H.~Li, ``Priority-based template matching intra
  prediction,'' in \emph{2008 IEEE International Conference on Multimedia and
  Expo}.\hskip 1em plus 0.5em minus 0.4em\relax IEEE, 2008, pp. 1117--1120.

\bibitem{zhang2015hybrid}
T.~Zhang, H.~Chen, M.-T. Sun, D.~Zhao, and W.~Gao, ``{Hybrid angular
  intra/template matching prediction for HEVC intra coding},'' in \emph{2015
  Visual Communications and Image Processing (VCIP)}.\hskip 1em plus 0.5em
  minus 0.4em\relax IEEE, 2015, pp. 1--4.

\bibitem{ibc}
C.~Pang, Y.~Wang, V.~Seregin, K.~Rapaka, M.~Karczewicz, X.~Xu, S.~Liu, S.~Lei,
  B.~Li, and J.~Xu, \emph{“Non-CE2: Intra block copy and inter signaling
  unification”}, JCTVC-T0227, 20th Meeting: Geneva, Feb 2015.

\bibitem{matsuo2009intra}
S.~Matsuo, S.~Takamura, and Y.~Yashima, ``Intra prediction with spatial
  gradients and multiple reference lines,'' in \emph{Picture Coding Symposium,
  2009. PCS 2009}.\hskip 1em plus 0.5em minus 0.4em\relax IEEE, 2009, pp. 1--4.

\bibitem{MultiLine}
J.~Li, B.~Li, J.~Xu, R.~Xiong, and G.~J. Sullivan, \emph{{Multiple line-based
  intra prediction}}, JVET-C0071, May 2016.

\bibitem{MultiLineITRI}
Y.~Chang, P.~Lin, C.~Lin, J.~Tu, and C.~Lin, \emph{{Arbitrary reference tier
  for intra directional modes}}, JVET-C0043, May 2016.

\bibitem{robertson2005dct}
M.~A. Robertson and R.~L. Stevenson, ``{DCT} quantization noise in compressed
  images,'' \emph{IEEE Transactions on Circuits and Systems for Video
  Technology}, vol.~15, no.~1, pp. 27--38, 2005.

\bibitem{norkin2012hevc}
A.~Norkin, G.~Bjontegaard, A.~Fuldseth, M.~Narroschke, M.~Ikeda, K.~Andersson,
  M.~Zhou, and G.~Van~der Auwera, ``{HEVC deblocking filter},'' \emph{IEEE
  Transactions on Circuits and Systems for Video Technology}, vol.~22, no.~12,
  pp. 1746--1754, 2012.

\bibitem{zhao2011fast}
L.~Zhao, L.~Zhang, S.~Ma, and D.~Zhao, ``Fast mode decision algorithm for intra
  prediction in hevc,'' in \emph{Visual Communications and Image Processing
  (VCIP), 2011 IEEE}.\hskip 1em plus 0.5em minus 0.4em\relax IEEE, 2011, pp.
  1--4.

\bibitem{hm}
\emph{https://hevc.hhi.fraunhofer.de/svn/svn\_HEVCSoftware/tags/HM-16.9/}.

\bibitem{ctc}
F.Bossen, \emph{“Common test conditions and software reference
  configurations”}, JCTVC-L1100, 12th Meeting: Geneva, Jan 2013.

\bibitem{JVETctc}
K.~Suehring and X.~Li, \emph{“JVET common test conditions and software
  reference configurations”}, JVET-B1010, 2nd Meeting: San Diego, Feb 2016.

\bibitem{psnr}
G.~Bjøntegaard, \emph{“Improvements of the BD-PSNR model”}, Document
  VCEG-AI11, July 2008.

\end{thebibliography}

}

\end{document}